\documentclass[lettersize,journal]{IEEEtran}

\usepackage{amsmath,amsfonts}
\usepackage{array}
\usepackage{textcomp}
\usepackage{stfloats}
\usepackage{verbatim}
\usepackage{graphicx}
\usepackage{cite}

\usepackage[frozencache=true,cachedir=minted-cache]{minted}

\usepackage{comment}
\usepackage[ruled,linesnumbered,vlined]{algorithm2e}

\usepackage{bbding}         
\usepackage{colortbl}       
\usepackage{threeparttable} 
\usepackage[normalem]{ulem}
\usepackage{cite}
\usepackage{xurl} 
\usepackage{amssymb}
\usepackage{booktabs}

\let\labelindent\relax
\usepackage{enumitem}

\usepackage{minted}
\usepackage{multirow}
\usepackage{multicol}
\usepackage{subcaption}
\usepackage{pifont}

\usepackage{scalerel}
\usepackage{tikz}
\usetikzlibrary{svg.path}

\definecolor{orcidlogocol}{HTML}{A6CE39}
\tikzset{
  orcidlogo/.pic={
    \fill[orcidlogocol] svg{M256,128c0,70.7-57.3,128-128,128C57.3,256,0,198.7,0,128C0,57.3,57.3,0,128,0C198.7,0,256,57.3,256,128z};
    \fill[white] svg{M86.3,186.2H70.9V79.1h15.4v48.4V186.2z}
                 svg{M108.9,79.1h41.6c39.6,0,57,28.3,57,53.6c0,27.5-21.5,53.6-56.8,53.6h-41.8V79.1z M124.3,172.4h24.5c34.9,0,42.9-26.5,42.9-39.7c0-21.5-13.7-39.7-43.7-39.7h-23.7V172.4z}
                 svg{M88.7,56.8c0,5.5-4.5,10.1-10.1,10.1c-5.6,0-10.1-4.6-10.1-10.1c0-5.6,4.5-10.1,10.1-10.1C84.2,46.7,88.7,51.3,88.7,56.8z};
  }
}

\newcommand\orcidicon[1]{\href{https://orcid.org/#1}{\mbox{\scalerel*{
\begin{tikzpicture}[yscale=-1,transform shape]
\pic{orcidlogo};
\end{tikzpicture}
}{|}}}}

\usepackage{hyperref} 
\usepackage{cleveref}

\crefformat{section}{\S#2#1#3}
\crefname{figure}{Figure}{Figures}
\crefname{table}{Table}{Tables}
\crefname{algorithm}{Algorithm}{Algorithms}

\setminted[c]{
	frame=lines,
	xleftmargin=1em,
	numbersep=0.2em,
	framesep=2mm,
	baselinestretch=1,
	fontsize=\footnotesize,
	tabsize=2,
	linenos,
	breaklines}
 
\newcommand{\jim}[1]{\textcolor{red}{[jim: #1]}}
\newcommand{\wenbo}[1]{\textcolor{blue}{Wenbo:#1}}
\newcommand{\code}[1]{\texttt{{\detokenize{#1}}}}
\newcommand{\zhiyun}[1]{\textcolor{cyan}{[\textbf{Zhiyun:} #1]}}

\newcommand{\jiayi}[1]{\textcolor{magenta}{[\textbf{Jiayi:} #1]}}

\newcommand{\PP}[1]{
\vspace{3px}
\noindent{\bf \IfEndWith{#1}{.}{#1}{#1.}}
}

\usepackage{tikz}
\newcommand*\emptycirc[1][1ex]{\tikz\draw (0,0) circle (#1);} 
\newcommand*\halfcirc[1][1ex]{%
	\begin{tikzpicture}
	\draw[fill] (0,0)-- (90:#1) arc (90:270:#1) -- cycle ;
	\draw (0,0) circle (#1);
	\end{tikzpicture}}
\newcommand*\fullcirc[1][1ex]{\tikz\fill (0,0) circle (#1);}

\newcommand{\name}{FSKO-AUTO}
\newcommand{\exploit}{PagePwn}

\definecolor{mygray}{gray}{0.9}

\begin{document}

\title{Beyond Control: Exploring Novel File System Objects for Data-Only Attacks on Linux Systems}

\author{Jinmeng Zhou, Jiayi Hu, Ziyue Pan, Jiaxun Zhu, Wenbo Shen, Guoren Li, Zhiyun Qian

\IEEEcompsocitemizethanks{\IEEEcompsocthanksitem 
Jinmeng Zhou, Jiayi Hu, Ziyue Pan, Jiaxun Zhu and Wenbo Shen are with the College of Computer Science and Technology at Zhejiang University, Hangzhou, Zhejiang, 310027, China.
Email: \{jinmengzhou, hujiayi, ziyuepan, sevenswords, shenwenbo\}@zju.edu.cn; \protect
\IEEEcompsocthanksitem Guoren Li and Zhiyun Qian are with the Department of Computer Science and Engineering, University of California, Riverside 92521, USA.
Email: gli076@ucr.edu and zhiyunq@cs.ucr.edu;
\protect
\IEEEcompsocthanksitem Wenbo Shen is the corresponding author. 
}}



\markboth{IEEE TRANSACTIONS ON INFORMATION FORENSICS AND SECURITY, VOL. XX, 20XX}%
{Shell \MakeLowercase{\textit{et al.}}: A Sample Article Using IEEEtran.cls for IEEE Journals}


\maketitle

\begin{abstract}
The widespread deployment of control-flow integrity has propelled non-control data attacks into the mainstream.
In the domain of OS kernel exploits, by corrupting critical non-control data, local attackers can directly gain root access or privilege escalation without hijacking the control flow.
As a result, OS kernels have been restricting the availability of such non-control data.
This forces attackers to continue to search for more exploitable non-control data in OS kernels.
However, discovering unknown non-control data can be daunting because they are often tied heavily to semantics and lack universal patterns.

We make two contributions in this paper: (1) discover critical non-control objects in the file subsystem and (2) analyze their exploitability.
This work represents the first study, with minimal domain knowledge, to semi-automatically discover and evaluate exploitable non-control data within the file subsystem of the Linux kernel. 
Our solution utilizes a custom analysis and testing framework
that statically and dynamically identifies promising candidate objects.
Furthermore, we categorize these discovered objects into types that are suitable for various exploit strategies, including a novel strategy necessary to overcome the defense that isolates many of these objects.
These objects have the advantage of being exploitable without requiring KASLR, thus making the exploits simpler and more reliable.
We use 18 real-world CVEs to evaluate the exploitability of the file system objects using various exploit strategies. 
We develop 10 end-to-end exploits using a subset of CVEs against the kernel with all state-of-the-art mitigations enabled.
\end{abstract}


\section{Introduction}


Control flow hijacking used to be the mainstream attack method to exploit memory errors, but it became difficult with the introduction of Control Flow Integrity (CFI).
As a result, data-only attacks have gained popularity~\cite{US22WANG2:online,shen2017defeating, WillsRoo38:online}. 
For example, in the Linux kernel, attackers target \code{modprobe_path}, where overwriting its value with an attacker-controlled executable path grants root privileges~\cite{modprobe-path}. There is no need to collect and stitch together ROP gadgets in such data-only attacks.
However, \code{modprobe_path} is a global variable. Overwriting it requires bypassing KASLR, which randomizes the address of all global variables. 
This extra complexity is a hurdle for exploits and their reliability. Alternatively, heap variables also contain critical data that can be leveraged to achieve privilege escalation.
They are easier to exploit for common vulnerabilities, such as out-of-bound (OOB) and use-after-free (UAF).
This is because vulnerabilities offer relative heap memory write, whereas KASLR does not help.
For instance, attackers can spray \code{struct cred} objects and use vulnerabilities to overwrite their \code{uid} fields to \code{0}, achieving the root privilege.

%



However, because of these objects' popularity, they are restricted nowadays. 
They are even protected by the hypervisor in Android and Apple XNU kernels~\cite{samsung1, samsung2, appleoss24:online}.
Because the well-known critical objects are specifically protected through advanced defenses, attackers no longer target them even if they manage to achieve an arbitrary write primitive.

To fill the gap, we seek to identify additional heap objects suitable for data-only attacks. 
Unfortunately, discovering unknown non-control data (heap objects or not) can be daunting because they are tied heavily to semantics (e.g., user id) and lack universal patterns.
In this paper, we perform a systematic investigation of objects associated with the file subsystem in the Linux kernel.
This is because file systems contain many critical files, e.g., \path{/etc/passwd}.  If an attacker can manipulate a file system into overwriting such a critical file, an attacker can achieve privilege escalation directly~\cite{CVE-2023-29383_etc_passwd}.

Few exploitable objects have been identified in the file subsystem of the Linux kernel~\cite{lin2022dirtycred}.
This is due to the lack of a systematic and automatic analysis of critical objects based on their semantics.
The only work that attempts to collect critical non-control data automatically in the Linux kernel is KENALI~\cite{song2016enforcing}.
However, there are two serious limitations of the work:
(1) 
its heuristic is overly generic and misses or falsely identifies many objects;
(2) it does not analyze the exploitability of identified objects.

Our solution is two-fold.  
First, we leverage minimal domain knowledge of the file subsystem in the Linux kernel to identify file system key objects (FSKO) with critical fields more accurately and completely.
Specifically, the process starts with a few anchor objects in abstract layers of the file subsystem. 
Then, it identifies additional related objects in other layers through tailored propagation rules.
We propose a framework named \name{} that integrates static analysis with dynamic testing as a strategy to identify FSKOs and verify that their semantics are appropriate for exploitation.
In total, \name{} finds 23 critical fields within heap objects whose corruption directly achieves privilege escalation.
Second, we map these objects into three targeted exploit strategies suitable for achieving privilege escalation using FSKO objects.
Since many FSKOs are located in isolated slab caches, we leverage a novel exploit strategy that uses ``bridge objects'' to construct a powerful and general primitive of page UAF, which can reliably exploit FSKOs without cross-cache attacks.
Out of 26 recent CVEs, our solution could demonstrate exploitability for 18 of them, implying its real-world impact and generality.
For a subset of the 18 CVEs, we managed to write 10 end-to-end exploits, achieving privilege escalation by corrupting the FSKO objects with various exploit strategies.
The results show that we can write diverse exploits without having to bypass KASLR, which makes the exploits simpler and more reliable than previous methods.

In summary, the paper makes three contributions.

\begin{itemize} [leftmargin=*]

    %

    \item \textbf{Semi-automatic FSKO identification and confirmation.} 
    With the help of static analysis and dynamic confirmation, we are able to drastically reduce the manual effort of identifying and confirming FSKOs. This led to 23 unique FSKO fields, covering all publicly exploited objects in the file subsystem.
    

    \item \textbf{Exploitability analysis and novel exploit strategy.} We analyze the exploitability of these FSKOs when paired with vulnerabilities of different capabilities and slab cache requirements. We propose a novel strategy that enables the many FSKOs to be exploitable, despite the fact that they are in isolated slab caches.

    \item \textbf{Practical evaluation and findings.} 
    We confirmed that 18 out of 26 CVEs are exploitable using the FSKOs and further developed 10 end-to-end exploits for a subset of them. The identified objects allow our exploits to avoid bypassing KASLRs, be future-proof against CFI, and advance the protection of well-known data. 



\end{itemize}

\section{Motivation and Threat Model}
\label{sec:overview}


\PP{Motivation}
After control flow attacks are mitigated, data-only attacks have gained popularity.
These attacks specifically target and corrupt a program's data rather than its control flow~\cite{BOP}. 
In the context of the Linux kernel, data-only attacks often target two types of data: (1) global variables and (2) heap variables. 
For global variables, the most widely targeted variable is \code{modprobe_path}. 
The variable is a string, and its corruption leads to the kernel executing an attacker-controlled executable file as root~\cite{modprobe-path}.
For heap variables, the most common ones are the \code{cred}~\cite{lin2022dirtycred} and the page table~\cite{dirty_pagetable}. Corrupting them can lead to privilege escalation directly.

\begin{figure}[H]
\centering
\includegraphics[width=0.35\textwidth]{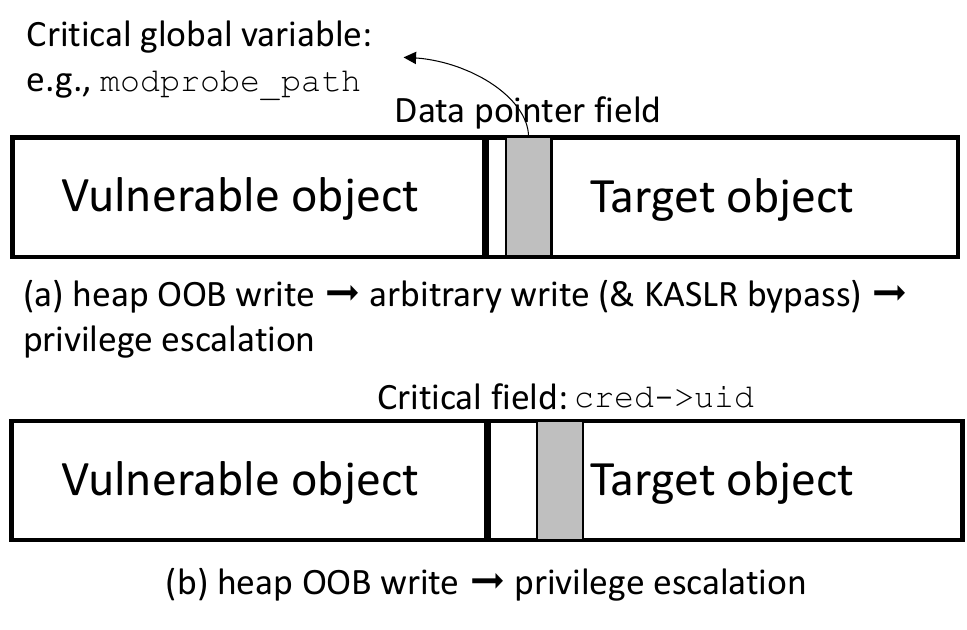}
\caption{Corrupting critical heap object vs. global variable.}
\label{fig:global-heap-attacks}
\vspace{-18pt}
\end{figure}

Using a common type of kernel vulnerability --- heap out-of-bound write --- as illustrated in ~\cref{fig:global-heap-attacks}, achieving privilege escalation by corrupting critical global variables is significantly more cumbersome than corrupting critical heap variables. As shown in ~\cref{fig:global-heap-attacks}(a), there are two extra steps: (1) deriving an arbitrary write primitive by corrupting a data pointer in a heap object and (2) bypassing KASLR to obtain the address of the global variable.
In contrast, \cref{fig:global-heap-attacks}(b) illustrates overwriting a critical heap variable directly lead to privilege escalation. 

The example assumes the heap variables can be co-located with the vulnerable object, which may not always be possible. This is especially the case when defenses increasingly target commonly exploited objects. First, isolation of \code{struct cred} in dedicated slab caches. It usually requires information leak to obtain their address and arbitrary write primitive to corrupt them. More importantly, commercial kernels apply advanced defenses, e.g., changing kernel configs and using hypervisors to deny illegitimate accesses, to prevent corruption of the critical objects in the dedicated slab cache~\cite{appleoss24:online, samsung1, samsung2}.
This motivates the discovery of additional critical heap objects.


\begin{figure*}[!th]
    \centering
    \includegraphics[width=.95\textwidth]{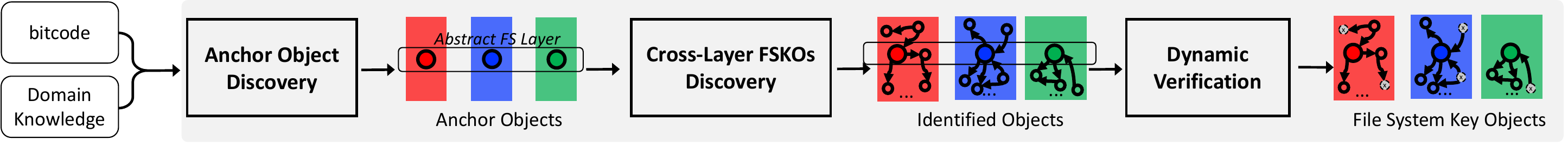}
    \caption{The work overview of identifying FSKOs spreading all layers in the file system.}
    \label{fig:overview}
    \vspace{-18pt}
\end{figure*}

\PP{Threat Model}
We assume an unprivileged user aims to achieve local privilege escalation by exploiting a memory error in the kernel. 
Specifically, the error allows limited memory writes (e.g., OOB write or UAF write), possibly as limited as a single-bit write.
We \textit{\textbf{do not}} assume the availability of arbitrary memory write primitives, i.e., write anywhere in memory. 
%
%


We assume the Linux kernel enables all modern mitigations, including CFI~\cite{linuxaddcfi:online}, W$\oplus$X~\cite{Extendin37:online}, KASLR~\cite{Kernelad62:online}, SMAP/SMEP (x86)~\cite{Supervis7:online}, and PAN/PXN (ARM). An unprivileged user cannot modify kernel code or inject code into the kernel data segment, and the kernel address is randomized. The kernel is restricted from accessing the data or executing the code in user space. Control flow cannot be hijacked. Besides, advanced defenses of well-known non-control data are safeguarded, including \code{modprobe_path}, \code{cred}, and page table~\cite{samsung1,samsung2}.

\PP{Attack Goal} 
Our goal is to achieve privilege escalation directly by corrupting non-control heap objects in the file subsystem of the Linux kernel. 
Two high-level approaches achieve the goal.
First, an attacker can corrupt certain heap objects (e.g., representing file permissions or owners) to gain write access to files that are read-only to the attacker~\cite{Arinerro42:online, DirtyCOW66:online, lin2022dirtycred}, e.g.,
\code{/etc/passwd}.
Second, an attacker can corrupt certain heap objects to turn non-setuid-root executable files into \emph{setuid root}~\cite{Privileg73:online}; the attacker can succeed when the executables are either controlled by the attacker or already has sufficient functionalities, e.g., \code{/usr/bin/vi}.
%
In this paper, we focus on exploring heap objects in the file system that match the above semantic descriptions in~\cref{sec:FSKO-identification} suitable for exploitation and evaluating the practical uses of them in~\cref{sec:exploit-analysis}.

\section{FSKO Identification}
\label{sec:FSKO-identification}

%
%
In this section, we focus on identifying objects in the file subsystem that can lead to privilege escalation when the fields inside specific objects are corrupted. 
We refer to them as \textbf{\textit{FSKOs}}, i.e., \underline{F}ile \underline{S}ystem \underline{K}ey \underline{O}bjects.
There are two main challenges in FSKO identification:

\textbf{\textit{Challenge 1}}: There are a large number of unique objects in the Linux kernel's file system, i.e., 3,553 structs and 554,161 fields in v5.14. Enumerating them manually is infeasible.
%

\textbf{\textit{Challenge 2}}: These objects are scattered across layers, encompassing both abstract layers (e.g., VFS) and concrete file system implementations (e.g., ext2, ext3). This makes it difficult to have a clean heuristic that works across layers.


To this end, we propose a new framework, \name{}, that combines static analysis and dynamic verification in a structured pipeline, as shown in ~\cref{fig:overview}.
For the first challenge, we propose to focus on identifying a small set of anchor objects with specific semantics and in abstract layers only to bootstrap the identification process. 
For the second challenge, from the anchor objects in abstract layers, we develop custom static analysis rules to search across other layers automatically.
Finally, we develop a dynamic verification process to confirm that statically identified objects are, in fact, with the right semantics and that their corruption leads to privilege escalation. 

\subsection{Anchor Object Discovery}
\label{sec:anchor-object}



We define \textit{\textbf{anchor objects}} to be FSKOs with fields that (1) hold the semantic that has security implications -- we consider specifically three classes of semantics, and (2) reside in the abstract layer, making it more manageable to analyze, compatible with different implementations.


\begin{figure}[!t]
    \centering
    \includegraphics[width=.8\linewidth]{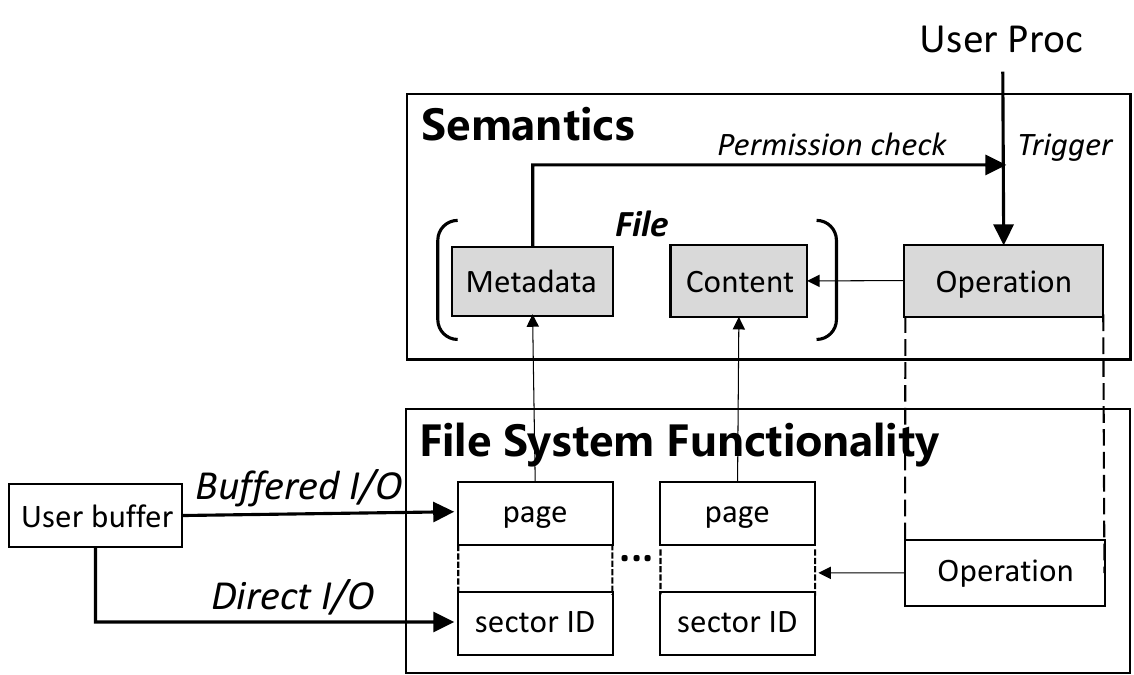}
    \caption{File system functionality in the abstract.}
    \label{fig:concept}
    \vspace{-18pt}
\end{figure}

\PP{Classes of Semantics.}
%
As shown in ~\cref{fig:concept}, file systems have a few basic design elements.
In general, a file consists of two parts: \emph{file content} that stores the actual data; \emph{file metadata} that encompasses properties like creation time, name, and permission. 
\emph{Operations} can be performed over files, e.g., read, write, and execute. 
Naturally, they correspond to the three classes of semantics that matter to privilege escalation. 
We consider this minimal domain knowledge necessary for identifying critical objects in the file subsystem.

\textit{(1) Metadata}:
Unprivileged users can corrupt permission or ownership objects to obtain authorization over files that are not owned by the attacker.
For example, an attacker can weaken the permission or change the ownership so that a sensitive read-only file is writable by the attacker~\cite{Arinerro42:online}.

\textit{(2) Content}: Corrupting content of sensitive files can lead to privilege escalation, e.g., \code{/etc/passwd}. In terms of heap objects, we know that file content can be represented as caches in memory~\cite{DirtyCOW66:online}, for buffered I/O. 
In the case of direct I/O, the write will occur directly on the hard drive.


\textit{(3) Operation:} An attacker can corrupt a file operation (represented by some heap objects) from a read or an execute into a write, resulting in the corruption of file content indirectly. This attack strategy is new and has not been observed in practice. 



\begin{figure}[!t]
\centering
\includegraphics[width=0.25\textwidth]{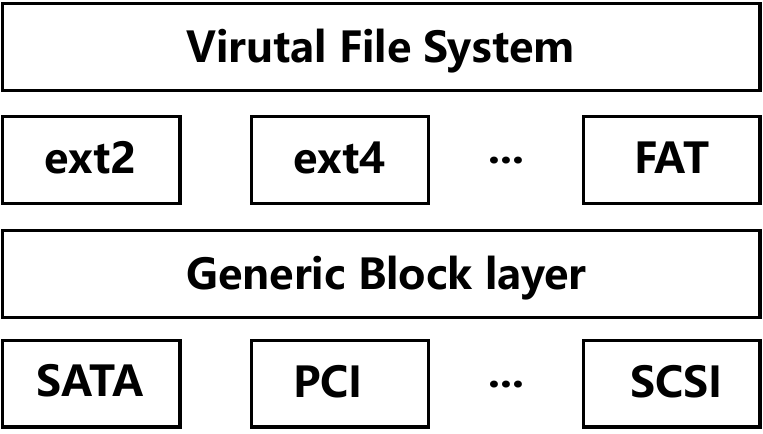}
\caption{Simplified file system layers in Linux kernel.}
\label{fig:simple_file_system}
\vspace{-18pt}
\end{figure}

\PP{Abstract File System Layers.}
Based on the domain knowledge, the Linux file subsystem is designed in layers, similar to any complex system.
As shown in \cref{fig:simple_file_system}, there are four layers from the top to the bottom: virtual file system (VFS), file system implementations, generic block, and drivers.
Specifically, we consider two well-known \textbf{\textit{abstract FS layers}} in the Linux kernel: the \textit{\textbf{VFS layer}} and the \textit{\textbf{generic block layer}}. The former sits immediately below the syscall interface, e.g., invoked by \code{open}, \code{read}, and \code{write}.
The latter sits below the layer of file system implementations that calculate file location on disk, so it connects the memory page and the disk sector. Both layers are agnostic to different underlying file system implementations (e.g., ext2, ext4) and various disk drivers.

We choose abstract layers over the concrete file system implementation layers because of the smaller and more manageable scope. Sensitive FSKOs must already be defined in the abstract layers and can be propagated into various concrete file system implementation layers, which we will capture automatically through static analysis.

\PP{Anchor Object Identification}
To identify objects relating to the three classes of semantics described earlier, we first identify the syscalls that can potentially affect the corresponding semantics (e.g., \code{open()} relates to metadata). We then review the functions involved in the syscalls that fall under the two abstract layers. For a specific class of semantic, we also follow certain heuristics to help us narrow down the search scope.





\textit{(1) Metadata}:
We rely on well-defined keywords and macros that identify the permission objects in the VFS layer.
Specifically, we grep variable names and fields with ``uid'' and ``gid'' that identify \code{inode->i_uid} and \code{inode->i_gid}
as the user ID and group ID of the file owner.
Regarding permission objects, we reuse macros from a previous work aiming to identify permission-check-related kernel functions~\cite{zhou2022automatic}. 
We limit our scope to macros related to discretionary access control (DAC).
Specifically, we include the three key macros: \code{MAY_EXEC}, \code{MAY_READ}, and \code{MAY_WRITE}, as well as \code{S_ISUID} and \code{S_ISGID} representing setuid and setgid permissions.
From these macros, we then statically track what variables are ``tainted'' by them using data flow analysis.

\textit{(2) Content}:
As alluded to earlier, file content can be represented in memory (buffered I/O) or on disk (direct I/O). Specifically, the in-memory caches are accessible via data pointers; the on-disk file content is addressable by sector numbers. 
In reviewing the syscall documentation, we realize that a flag in the \code{open()} syscall controls the access mode, and therefore, we consider the access mode separately for \code{read()} and \code{write()} syscalls, i.e., we review their call graphs conditioned on the access mode.
In summary, we find page caches accessed through \code{inode->i_mapping} and \code{file->f_mapping} in the VFS layer, which is one anchor object. We find an anchor object in the generic block layer that contains the sector numbers and interacts with the disk, \code{bio->bi_iter->bi_sector}.

\textit{(3) Operation:}
%
We primarily focus on reviewing the \code{read()} and \code{write()} syscalls as the file operation is likely encoded in some form internally in the two abstract layers. 
To be more efficient, we compare the call graphs of the two syscalls, focus on the functions that are shared by both, and stop when the call graphs diverge. The intuition is that the divergence should exist before the divergence that leads to the separate functions being invoked.
For buffered I/O, we find that there is no object representing file operations as the kernel quickly diverges by calling \code{vfs_read()} and \code{vfs_write()} to distinguish the read and write operations. 
On the other hand, for direct I/O, we find there is an object called \code{bio->bi_opf} in the generic block layer.

\subsection{Cross-Layer FSKOs Discovery}
\label{sec:type-based-object-discovery}
We develop custom static analysis rules that track the propagation of anchor objects into other concrete implementation layers. 
In order to track the flow of these objects properly, these rules operate forward and backward. 
Ultimately, the key is to identify other objects that carry the same or similar semantics.
Our static analysis takes critical fields identified in anchor objects as input and outputs the derived fields in other objects.
At a high level, the rules explore the data dependency and control dependency of the anchor objects with specific considerations of file system characteristics.
The static analysis consists of 3K LoC using LLVM/Clang 14.0.0.

\PP{Definitions}
%
%
We categorize file metadata, content, and operation into two types of data, \textit{flags} and \textit{references}, to simplify the description of our algorithm.
\textit{Flags} represent file metadata and operations; overwriting the corresponding fields of such objects or object fields with constants (permission of 777 allows anyone to read/write a file) can directly lead to privilege escalation.
%
\textit{References} represent file content, i.e., file cache (pointers) and disk locations (sector numbers), which correspond to buffered I/O and direct I/O, respectively.
Modifying the reference from pointing to a regular writable file (low-privilege) to a read-only (high-privilege) file can achieve root.

\begin{table}[t]
\centering
\caption{Custom Propagation Rules.}
\vspace{-5pt}
\label{tab:rules}
\resizebox{1\linewidth}{!}{
\begin{tabular}{l|p{3.2cm}|p{3cm}}

\toprule[0.1pt]
\toprule[0.1pt]
        &     Data dependency                            &    Control dependency   \\ \hline
Flag      &  Assignment (optionally involving logical ops)      &  Value influence by control \\ \hline

Reference &  Assignment (optionally involving arithmetic ops)   &  Reference assignment influenced by a flag variable \\ 

\bottomrule[0.1pt]
\bottomrule[0.1pt]
\end{tabular}}

\vspace{-10pt}
\end{table}

\PP{Analysis} 
First of all, to analyze a complex subsystem like the file system, we already face a significant challenge of pointer analysis, which is a foundation of data flow analysis.
This is because multiple syscalls are dependent on each other, e.g., \code{open()}, \code{read()}, \code{write()}, and \code{mmap()}.
One syscall may store something on the heap, and another syscall may retrieve it subsequently (i.e., aliases across syscalls). 
We borrow the technique proposed in prior work, which uses access-path-based approaches (considering multiple levels of types) to identify aliases across syscalls~\cite{zhang2021statically}. In addition, we use the state-of-the-art indirect call analysis~\cite{lu2023practical} that prunes many false indirect call edges.
We summarize the rules for tracking flags and references in~\cref{tab:rules}. 

\PP{Rule 1: Flag Propagation}
%
%
%
When tracking a flag (e.g., a field within an object), it is considered to potentially influence or be influenced by another variable through data (i.e., assignments) or control dependency (i.e., conditional branches).
Specifically, we consider both forward and backward dependencies.
We customize the data dependency by tracing the direct assignment (e.g., \code{a = b}) and the assignment involving logical expressions (e.g., \code{a = b & STATE}), including AND(\&), OR(\code{|}), NOT($\sim$). This is based on the observation that flags typically involve bitwise operations. Arithmetic operations are not included because flags are unlikely to be used in addition or subtraction operations. 
This helps us eliminate false positives. 

%


\PP{Rule 2: Reference Propagation}
The rules tracking a reference variable are also shown in~\cref{tab:rules}.
First, we consider a reference object propagates its particular value to another reference through solely data dependency (an exception will be discussed later). If other variables also store sector numbers and \code{page} pointers, e.g., due to assignments, these variables will be considered references.
We also customize the data dependency rules to allow arithmetic operations because a reference can add a relative value to find a neighbouring reference. This customization excludes other types of data dependency, such as bitwise operations in logical expressions. 
%


We consider control dependencies in a limited capacity: \emph{flag} variables can determine a reference's value through conditional branches (e.g., whether a reference will point to a page cache in DirtyCow~\cite{DirtyCOW66:online}).
We generalize this pattern to consider the flags in a condition where (1) at least one of the branches sees an assignment to a reference, (2) the assignment differs in the two branches -- either one branch lacks an assignment or the reference is assigned different values. 




\subsection{Dynamic Verification}
\label{sec:dynamic_verification}
After identifying all candidate FSKOs and their specific fields, \name{} dynamically them by forcefully corrupting their values at runtime and observing the effects on privilege escalation.
Dynamic verification is necessary because we observe some objects seem to contain critical fields but cannot lead to privilege escalation. 
For example, paths may be infeasible or require privileges to reach.
For example, our static analysis finds that \code{iattr->ia_mode} is assigned to \code{inode->i_mode} where the latter field represents the file permission. This makes the former object seemingly exploitable since we can manipulate the permission. However, only the file's owner can trigger the assignment when setting the file attributes, making the object non-exploitable.

\begin{figure*}[!t]
    \centering
    \includegraphics[width=.85\linewidth]{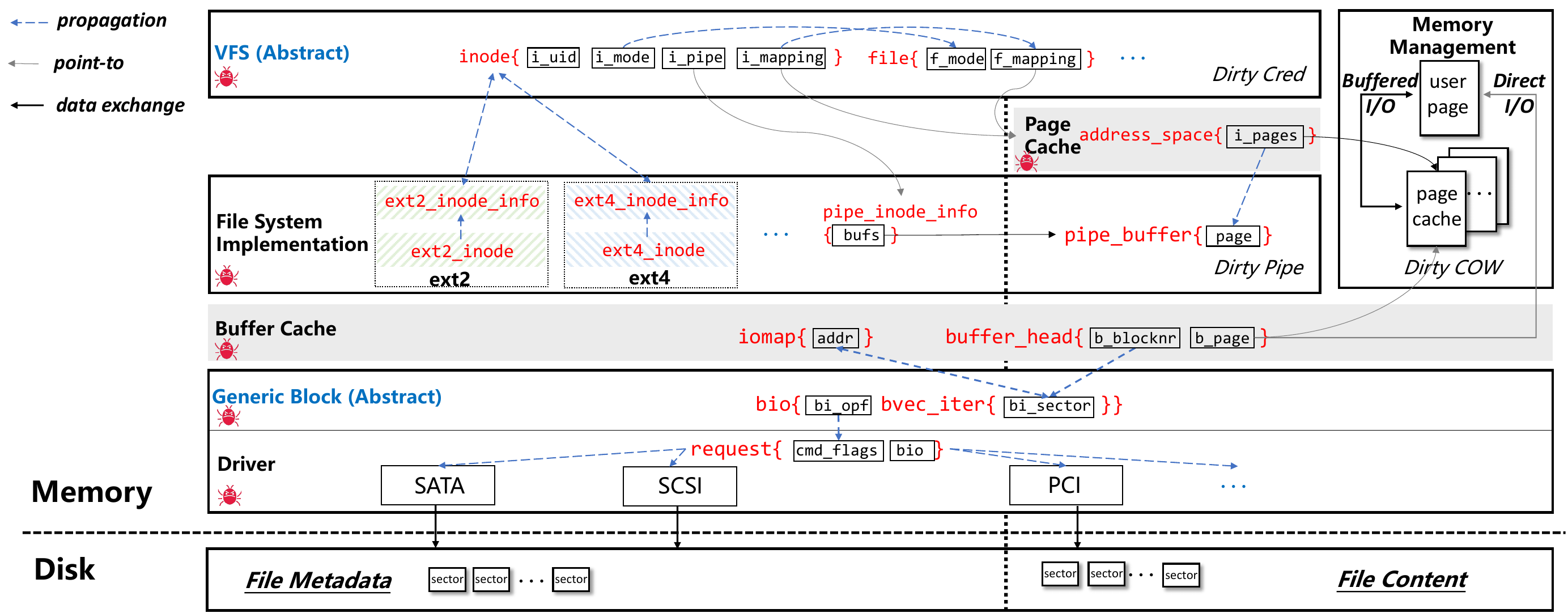}
    \caption{Some key objects identified by \name{} are highlighted in red, propagating all file system layers.}
    \label{fig:file-system}
    \vspace{-18pt}
\end{figure*}



We develop a process for dynamic verification.
First, we collect userspace test cases to cover file-related operations such as read, write, and mmap. 
We rely on two sources: (1) syzkaller traces collected during our fuzzing campaign and (2) manually curated test cases. We need to curate test cases because the vanilla syzkaller is insufficient. For example, syzkaller does not create multiple file systems to exercise the data structures specific to each file system (e.g., ext2, ext3). Fortunately, the number of file system-related syscalls is limited, and we managed to cover more than half of FSKO field candidates in the end (see details in \S\ref{sec:imp-eva}).

Next, for specific FSKO fields covered by test cases at runtime, we simulate a successful attack by corrupting these fields. We instrument the kernel to preserve a certain memory to load (record) and store (corrupt) value for each FSKO using an LLVM pass.
At first glance, it is challenging to simulate a successful attack because we do not know what value to write into a specific field of an FSKO.
To overcome this issue, we apply differential analysis to automatically retrieve the appropriate value that can likely lead to privilege escalation. 
We apply this strategy to the two conceived attack approaches described in the attack goals of \S\ref{sec:overview}. 
Using the approach of ``turning a read-only file into a writable file'' as an example, below is how we perform the differential testing:

(1) Record the value of a specific FSKO field of interest (e.g., \code{file->f_mode}) in a successful write operation (against a test file we create).
(2) Replace (corrupt) the value of the same FSKO field in a read operation (e.g., against \path{/etc/passwd}) with the value recorded in the write operation. 
In the case of \code{file->f_mode}, this effectively changes the file permission from read-only to writable.
(3) Observe whether the read operation is turned into a successful write operation. If so, the FSKO field is deemed a valid one.

\subsection{Summary of Identified Objects}

\name{} initially identified 258 candidate FSKOs with static analysis.
After we apply dynamic verification, 25 heap FSKOs and 8 stack FSKOs are confirmed to be effective for privilege escalation, we focus on the heap objects in this paper and will discuss them in more detail in \S\ref{sec:imp-eva}.

In \cref{fig:file-system}, we illustrate and categorize some representative objects that are identified across various layers of the Linux kernel file subsystem, from the VFS layer at the top all the way to the driver layer at the bottom.
We can see the left part represents the metadata, while the right side represents the content, with a dotted vertical line separating them. 
Taking \code{inode->i_mode} as an example anchor representing file permission, it propagated to another object \code{file->f_mode} which represents permission as well. 
Of course, to corrupt these objects, we need to first be able to allocate them. 
For \code{struct inode}, we find it is possible to allocate it individually and as part of bigger objects, e.g., \code{ext2_inode_info}, 
that are specific to file system implementations.
%
Ultimately, these newly discovered objects give the attack significantly more flexibility in choosing heap objects.
We will discuss a more complete list of FSKOs in~\cref{tab:identified-flag}. 

It is worth noting that all key objects leveraged in prior exploits (which also target file system-related objects), including DirtyPipe, DirtyCow, and DirtyCred, are discovered in our work.
For example, we find \code{pipe_buffer->flags} and \code{vm_fault->flags} which are used by DirtyPipe and DirtyCOW to manipulate flags that control which reference object is used.
We also discovered \code{file->f_mapping} which represents a file page.
%
DirtyCred targeted the objects containing \code{struct file}, which can be easily identified after pinpointing the basic fields representing file content. 

\begin{figure*}[!th]
    \centering
    \includegraphics[width=.95\textwidth]{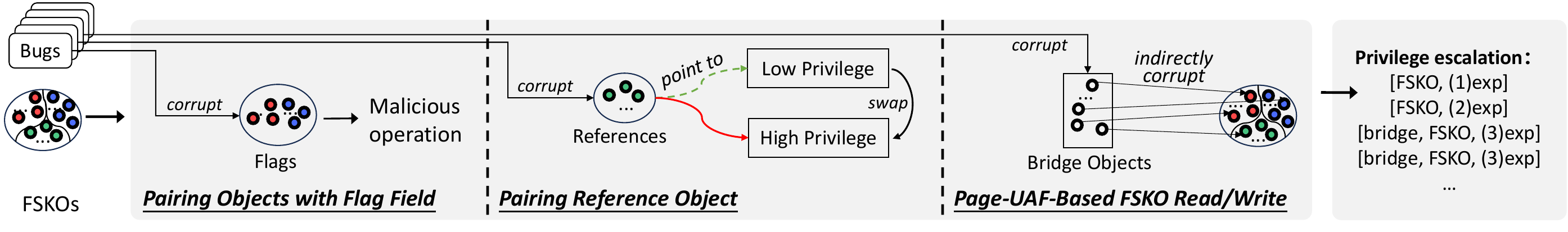}
    \vspace{-5pt}
    \caption{The work overview of exploitability analysis to find effective objects.}
    \label{fig:exploitability}
    \vspace{-18pt}
\end{figure*}

\section{Exploitablity Analysis}
\label{sec:exploit-analysis}

So far, we have gathered a list of FSKOs with specific critical fields that, when corrupted, lead to privilege escalation.
The dynamic verification step only verified the semantics of FSKO fields by forcefully overwriting them.
In practice, we need to pair these objects with specific vulnerabilities.
Generally, the more FSKOs, the higher the likelihood of successfully pairing one with a given vulnerability. We provide the exploitability analysis overview in~\cref{fig:exploitability}, including flag and reference corruption. Besides, we propose a new page-UAF strategy to pivot OOB/UAF bugs into a page UAF, allowing subsequent corruption of both flag and reference.

\PP{Pairing FSKOs with vulnerabilities}
There are two key dimensions related to pairing objects with vulnerabilities:

\noindent\underline{(1) Capability of a vulnerability.} As studied in ~\cite{chen2020koobe}, different vulnerabilities can exhibit different write capabilities, e.g., offset, length, and the possible values of the write (see \cref{fig:global-heap-attacks}). 
Depending on the specific capability, we might choose to pair an FSKO with fields at the right offset.

\noindent\underline{(2) Slab cache of the vulnerable object.}
In heap exploits, a target object (e.g., FSKO) needs to be placed in the same slab cache (e.g., \code{kmalloc-512}) that the vulnerable object resides in~\cite{chen2019slake}. 
The more FSKOs, the more likely we can find an object that co-locates with the vulnerable object. 
Otherwise, an attacker would resort to cross-cache attacks, which is considered much less reliable~\cite{autoslab}.

\begin{table}[t]
\centering
\renewcommand{\arraystretch}{1.1}
\caption{The requirements to corrupt FSKOs.}
\vspace{-5pt}
\label{tab:summary}
\resizebox{.9\linewidth}{!}{
\begin{tabular}{ccc}
\toprule[0.1pt]
\toprule[0.1pt]
 \textbf{Object Types} & \textbf{\begin{tabular}[c]{@{}c@{}}Information \\ Leakage\end{tabular}} & \textbf{\begin{tabular}[c]{@{}c@{}}Memory Corruption \\ Capability\end{tabular}} \\
\midrule
 Flags & / & Limited write    \\ \hline

 File cache pointers & Required  & Write controllable value \\ \hline

  File cache pointers  & /  & Write last few bits/bytes   \\ \hline
  Disk sector numbers  & /  & Write controllable value  \\ \hline

\bottomrule[0.1pt]

\end{tabular}}
\vspace{-10pt}
\end{table}

\subsection{Pairing Requirement \#1: Write Capability}

We summarize the write capability requirement of different types of objects in ~\cref{tab:summary}.

\PP{Pairing Flag Objects}
Corrupting objects with flag fields is straightforward. As long as they are overwritten as pre-determined constant values, privilege escalation will succeed. As mentioned in \S\ref{sec:dynamic_verification}, we can retrieve the target values at runtime and reuse them without information leak. We list the exact constant values expected for flags later in \cref{tab:identified-flag}.
The only requirement is that the vulnerability can overwrite at the specific field offset with an expected value.
In summary, \textit{flag} objects are generally easy to pair, requiring only limited write capabilities, e.g., setting a specific bit or writing a few bytes with constants or attacker-controlled values.



\PP{Pairing Reference Objects}
As mentioned earlier, reference objects include data that represent file content,
i.e., file page cache pointers or sector numbers.
Generally speaking, we envision the proper exploit strategy to be swapping the references so that they point to a sensitive file for corruption.

To overwrite page cache pointers, a straightforward approach is to swap to the sensitive file's page cache. However, it requires an extra step of information leaks and the corruption value being controlled by the attacker.
Alternatively, one can spray many file pages corresponding to sensitive files to make them co-located with the file page of an attacker-owned file. Overwriting the lower bits of the file page pointer (e.g., setting the lower two bytes to 0) will cause it to point to a nearby page. This method does not require information leak and imposes less stringent demands on the write capability, which has been used in prior work~\cite{lin2022dirtycred}. 
However, it needs to place the page caches of the low-privilege and high-privilege files adjacent to each other. This highly requires manipulation of the memory layout, which is difficult and unstable. 



To overwrite sector numbers, we find that it does not require any information leak because unprivileged users can obtain the sector numbers of any file through an \code{ioctl()} syscall with \code{FS_IOC_FIEMAP}~\cite{fiemap}.
This method only requires a write capability with attacker-controlled values.

\subsection{Pairing Requirement \#2: Slab Cache}

Many FSKOs are unfortunately located in dedicated slab caches, e.g., \code{filp_cachep} that hosts only a single type of objects. This makes them unlikely to be co-located with a vulnerable object.
To overcome this, we come up with a new and general exploit strategy that can indirectly achieve complete FSKO read and write without resorting to cross-cache attacks, called \textit{\textbf{Page UAF strategy}}. 

\vspace{3pt}
\noindent\underline{1. Page-level UAF and its Advantages.}
Given an initial primitive in the form of a constrained offset write, e.g., triggering an OOB, UAF, and Double Free (DF) vulnerability, we aim to construct and leverage a new and powerful primitive --- page-level UAF primitive. 
A page-level UAF represents a primitive where the attack can access a freed 4KB physical page (read/write) through a special dangling pointer. The freed page will be recycled and possibly used by objects in standard or dedicated slab caches. Through this primitive, an attacker gains complete read and write access to such objects (including FSKOs in dedicated slab caches). 
This strong primitive offers several advantages.
First, it avoids the complex cross-cache attacks and is thus much more reliable.
Second, the ability to corrupt such critical objects directly means that there is no need to bypass KASLR.
Third, no need for complex heap feng shui, making exploits more stable.



\vspace{3pt}
\noindent\underline{2. Page-level UAF construction.}
To create a page-level UAF, the key is to cause a UAF of the \code{struct page} objects. This is because a \code{struct page} (64 bytes, 0x40) object in the Linux kernel corresponds directly to a 4KB physical page. \emph{Freeing} a \code{struct page} object causes the corresponding 4KB page to be \emph{freed} as well.

\begin{figure}[!t]
  \centering
  \includegraphics[width=\linewidth]{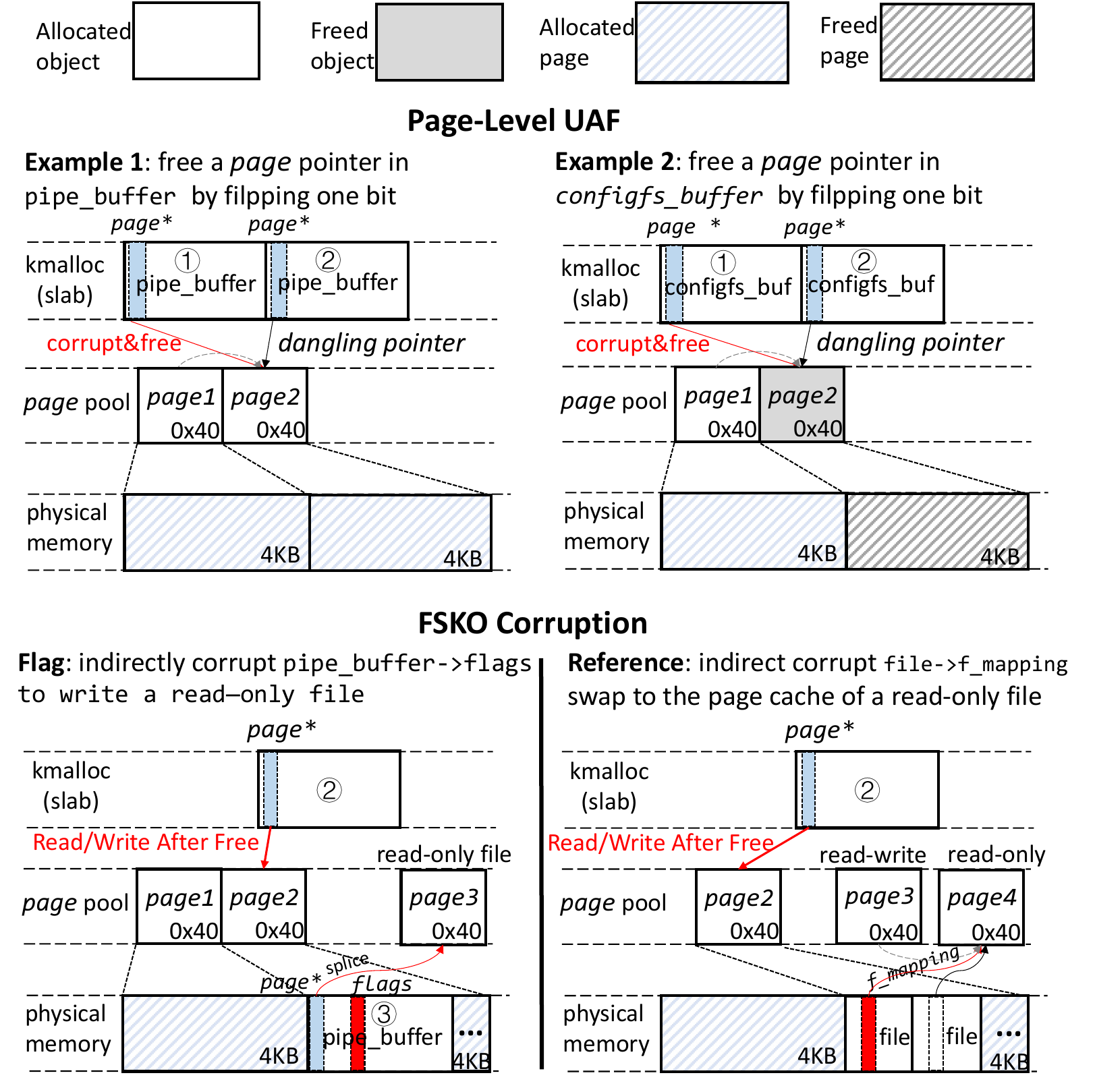}
  \vspace{-10pt}
  \caption{Use page-UAF strategy to corrupt flag and reference object, resulting in writing read-only files.}
  \label{fig:exploit}
  \vspace{-20pt}
\end{figure}

To construct the UAF of \code{struct page} objects, we need to leverage objects that contain pointers pointing to such \code{page} objects. Preferably, such objects should be allocated in standard slab caches (e.g., \code{kmalloc-512}) so that they can be paired with initial memory corruption primitives. 
More specifically, the idea is to corrupt the pointer field of bridge objects, so that two \code{page} pointers will point to the same \code{page} object, which can lead to a successful UAF~\cite{lin2022dirtycred}. 
Because bridge objects are located in standard slab caches, they can be sprayed and paired with initial memory corruption primitives such as OOB and UAF. 
As shown in ~\cref{fig:exploit}, 
once the \code{page} pointer in one bridge object is corrupted, it can be directed to another adjacent \code{page} object, which can trigger an invalid \code{page} free and lead to a dangling \code{page} pointer. 


%


\emph{Pivoting OOB \& UAF.}
We use write primitive to corrupt
the \code{page} pointer field in a bridge object (e.g., fields of \code{pipe_buffer} and \code{configfs_buf}) such that it points to another 64 byte (0x40) \code{page} object nearby. Specifically, if we corrupt/zero a page pointer’s last byte (in bridge object) with 0x00, it will point to another nearby struct page (when the byte is originally 0x40/0x80/0xc0). The pointer remains unchanged if the lowest byte is originally 0x00, which means the attackers corrupt nothing, and the kernel will not crash. This zeroing last few bits is a general exploitation step, and effectively causes two pointers to point to the same object~\cite{lin2022dirtycred}. A user-space program can trigger \code{free_pages()} on one of the objects, e.g., by calling \code{close()}, which will create a dangling pointer to the freed \code{page} object and the corresponding physical page.
In other words, we can read/write the corresponding physical page that is now considered freed by the OS kernel.
For example, one can write to a pipe, which will lead to a write of the physical page, via the \code{pipe_buffer} object.

\emph{Pivoting double free.}
For bugs that have double free primitives, which often can be achieved from UAF by triggering the \code{free()} operation twice. For the first free, we spray a harmless object (e.g., \code{msg_msg}) to take the freed slot, which takes attacker-controlled value from user space. Then, we trigger the kernel code to write the harmless object until reaching a certain offset, and use the FUSE technique~\cite{fuse_exploit} to stop the writing 
right before a planned offset -- corresponding to the page pointer field of a planned bridge object. Now, we trigger the free for a second time to spray the planned bridge object to take the slot. The writing process is restarted to continue overwriting the lower bits of the page pointer field, which leads to a page UAF.
Previously, to trigger page-level frees from double frees, one had to free an entire slab and then do cross-cache technique~\cite{lin2022dirtycred} whereas we do not need to.


\emph{Page adjacency.} 
It is clear that if we spray a sufficient number of bridge objects, they will be adjacent, as shown in Figure~\ref{fig:exploit}. However, we also need to ensure that the \code{struct page} objects (0x40-byte long) that bridge objects point to are also adjacent. Otherwise, corrupting the lower bits of a pointer in one bridge object may not cause it to point to a nearby \code{struct page} object.
Fortunately, this is easily achievable.
First, allocations for bridge objects and \code{struct page} objects are separate. For example, we can first spray \code{pipe_buffer} objects without allocating the corresponding pages. This allows us to spray \code{struct page} objects separately and make them contiguous.
Second, all \code{struct page} objects are maintained in a dedicated memory region so that they will not be mixed with other object types.
Third, Allocating physically contiguous pages ensures contiguous \code{struct page} objects.
We ensure \code{struct page} objects corresponding to nearby \code{pipe_buffer} are contiguous by forcing physical-contiguous pages to be allocated, i.e., after exhausting lower-order physical pages.
Besides, we can read the content of the freed-then-reused page through bridge objects. So even if we accidentally corrupt the \code{struct page} pointer to point to a wrong page, we can detect it and not trigger write to it, avoiding crashes.

\vspace{3pt}
\noindent\underline{3. FSKO corruption through page-level UAF.}
Now that we have a freed physical page and a dangling pointer that can read/write it. It is fairly easy to then allocate and corrupt FSKO objects.
This is because the flag and reference objects are eventually allocated through the buddy allocator at the page granularity.
Therefore, attackers can spay FSKOs into the freed page, as shown in~\cref{fig:exploit}, as long as they have already exhausted all existing slab caches.

~\cref{fig:exploit} showed two example FSKOs that we successfully exploited.
In the first example, we allocate \code{struct pipe_buffer} objects (they are both bridge objects and FSKOs). 
We then connect the pipe to a read-only file via \code{splice()} syscall, and achieve write permission by corrupting \code{pipe_buffer->flags}.
In the second example, we allocate two \code{struct file} objects.
The first object corresponds to a read-only file, and we then read its
\code{file->f_mapping}, which points to a data structure that represents the files's page cache (file content). We then allocate a second \code{struct file} object that corresponds to an attacker-created file. Next, we write its \code{f_mapping} field with the value of the first \code{file} object. This allows the attacker to use the second \code{file} to write into the first.

\textbf{Bridge object discovery.}
We query \code{structs} containing \code{struct page} fields and record those found in our fuzzing campaign and manually curated test cases.
We exclude the objects where a user cannot perform read and write operations over the page.
Finally, we list the exploitable bridge objects in \cref{tab:identified-page}.
8 objects are elastic, i.e.,  with variable sizes, making them ideal for satisfying the slab cache requirement.

To our knowledge, we have not seen widespread use of bridge objects to achieve page-level UAF in real-world exploits. The only example we are aware of is a CTF competition~\cite{arttnba30:online} that uses the \code{struct pipe_buffer} as a bridge object. However, we note that \code{struct pipe_buffer} is being isolated into \code{kmalloc-cg} slab using flag \code{GFP_ACCOUNT} after Linux kernel v5.14, which requires cross-cache corruption in future
exploits. Systematically identifying more bridge objects in standard slab caches is necessary since it provides more possibilities to match vulnerabilities with different capabilities~\cite{chen2020koobe}.


\textbf{Summary.}
The core of the page UAF strategy leverages insufficient page isolation. Unlike the slab caches with over a hundred isolated groups which offer strong isolation (as is the case for most FSKOs), there are only at most six page groups that are isolated from each other~\cite{migratetype}. Because of this, the pages whose content is fully controlled by users (e.g., pipe pages) are not isolated from the page used in the slab. In other words, the slabs (that host FSKOs) can reuse the freed user-controlled pages, rendering the page-UAF strategy successful. This strategy also works for not just FSKOs but any other types of objects in isolated slab caches. 
\section{Evaluation}
\label{sec:imp-eva}

\textbf{Experiment setup.}
(1) FSKO identification:
We performed our static analysis and dynamic testing against the Linux kernel v5.14 and v6.6. Choosing two versions is to test the generality of our tool and methodology based on stable FS conventions. We verified that all effective heap objects exist in both versions because they are related to core FS semantics and do not change frequently. The v5.x series is still widely used in real-world environments, and we choose 5.14 as a representative, besides, the v6.6 is the latest stable version.
(2) Exploitability analysis: 
We sample 26 recent Linux kernel CVEs that are either OOB, UAF, or double-free from 2020 to 2023. 
This includes 24 vulnerabilities from prior work~\cite{lin2022dirtycred},
and 2 additional OOB vulnerabilities missed by the prior work.

\begin{table}[t]
\centering
\renewcommand{\arraystretch}{1.1}
\caption{Statistical results of identified objects.}
\label{tab:count}
\vspace{-5pt}
\resizebox{.75\linewidth}{!}{
\begin{tabular}{lllrrr}

\toprule[0.1pt]
\toprule[0.1pt]
\multirow{2}{*}{} & \multirow{2}{*}{Category} & \multirow{2}{*}{\#} & \multicolumn{3}{c}{Security Impact} \\\cline{4-6}
            &               &       & PE        & NPE       & No test        \\ \hline
            & Metadata              &  148   &  9       &  61       & 78 \\
Flag        & Operation             &  32   &  8       &  12        & 12\\
            & COW                   &  5   &  2       &  0       & 3\\ \hline
\multirow{2}{*}{Reference} & Sector &  75   &  10       &   36      & 29\\
            & Page Cache            &  22   &  4       &  3       & 15 \\ \hline
 Sum     & /                        & 282      & 33        & 112   & 137\\

\bottomrule[0.1pt]
\bottomrule[0.1pt]
\end{tabular}}

\begin{tablenotes} \scriptsize

\item PE: can lead to Privilege Escalation after dynamic validation.
\item NPE: cannot lead to Privilege Escalation after dynamic validation.
\item No test: cannot dynamically verify.
\end{tablenotes}
\vspace{-15pt}
\end{table}

\begin{table}[t]
\centering
\renewcommand{\arraystretch}{1.1}
\caption{Identified FSKOs. The attack succeeds (\CheckmarkBold{}) as long as corrupting them with the targeted values. }
\vspace{-5pt}
\label{tab:identified-flag}
\resizebox{1\linewidth}{!}{
\begin{tabular}{c|c|c|c}
\toprule[0.1pt]
\toprule[0.1pt]
 Data   &  Category     &  Target value       & Memory Cache \\ \hline

\rowcolor{mygray}
\multicolumn{1}{c|}{\begin{tabular}[c]{@{}c@{}}\code{ext4_inode_info}\\\code{->vfs_inode->i_uid->val}\end{tabular}}
& metadata (A)    & \code{user id}    &  ext4\_inode\_cachep \CheckmarkBold           \\

\multicolumn{1}{c|}{\begin{tabular}[c]{@{}c@{}}\code{posix_acl}\\\code{->a_entries->e_perm}\end{tabular}}   & metadata (A)   & \code{MAY_WRITE}    & variable size \CheckmarkBold           \\

\rowcolor{mygray}
\multicolumn{1}{c|}{\begin{tabular}[c]{@{}c@{}}\code{ext4_inode_info}\\\code{->vfs_inode->i_mode}\end{tabular}}   & metadata (A)    & \code{S_ISUID, S_IWOTH}    & ext4\_inode\_cachep \CheckmarkBold           \\

\code{ext4_inode->i_uid}    & metadata   & \code{user id}    & page \CheckmarkBold           \\

\rowcolor{mygray}
\code{ext4_inode->i_mode}    & metadata   & \code{S_ISUID,S_IWOTH}    & page \CheckmarkBold           \\




\code{file->f_mode}     & metadata  & \code{FMODE_WRITE}       & filp\_cachep \CheckmarkBold           \\

\rowcolor{mygray}
\code{iattr->ia_mode}   & metadata  & \code{S_ISUID,S_IWOTH}       & \XSolidBrush           \\

\multicolumn{1}{c|}{\begin{tabular}[c]{@{}c@{}}\code{vm_area_struct}\\\code{->vm_flags}\end{tabular}}
 & metadata  & \code{VM_WRITE}     & vm\_area\_cachep \CheckmarkBold    \\

\rowcolor{mygray} 
\code{nfs_pgio_header->rw_mode}  & metadata  & \code{FMODE_WRITE}   &  
\XSolidBrush \\

\code{nfs_fattr->mode}  & metadata  & \textbf{---}   &\textbf{---}             \\

\rowcolor{mygray}
\code{nfs_open_context->mode} & metadata  & \code{FMODE_WRITE}  &\XSolidBrush             \\

\code{nfs_openargs->mode}   & metadata   & \code{FMODE_WRITE}    & \XSolidBrush         \\

\rowcolor{mygray}

\multicolumn{1}{c|}{\begin{tabular}[c]{@{}c@{}}\code{pnfs_layout_range}\\\code{->iomode}\end{tabular}} & metadata   & \code{IOMODE_RW}     & \textbf{---}         \\








\code{xfs_buf->b_flags} & operation & \code{XBF_WRITE} & \XSolidBrush \\



\rowcolor{mygray}
\code{iomap_dio->flags}  & operation     & \code{IOMAP_DIO_WRITE}  & kmalloc-96 \CheckmarkBold  \\

\code{dio->op}   &  operation  & \code{REQ_OP_WRITE}   & dio\_cache \CheckmarkBold   \\

\rowcolor{mygray}

\code{request->cmd_flags}  & operation   &  \code{REQ_OP_WRITE}  & kmalloc; driver-specific \CheckmarkBold   \\

\code{bio->bi_opf} & operation (A)  &  \code{REQ_OP_WRITE}   & \multicolumn{1}{c}{\begin{tabular}[c]{@{}c@{}}slab cache;\\driver-specific \CheckmarkBold\end{tabular}}  \\ 

 \rowcolor{mygray}
\code{scsi_cmnd->cmnd[0]} & operation &  \code{WRITE_32}   & sd\_cdb\_pool \CheckmarkBold     \\

\code{aio_kiocb->kiocb.flags} & operation  & \code{IOCB_WRITE} & \textbf{---}  \\

\rowcolor{mygray}
\code{dm_io_request->bi_op}   & operation & \code{REQ_OP_WRITE}     & \XSolidBrush   \\


\code{pipe_buffer->flags} & COW  & \multicolumn{1}{c|}{\begin{tabular}[c]{@{}c@{}}\code{PIPE_BUF_FLAG_}\\\code{CAN_MERGE}\end{tabular}}   & variable size \CheckmarkBold \\

\rowcolor{mygray}
\code{fuse_copy_state->write} & COW  & \code{0} & \textbf{---} \\






\code{bio->bi_iter->bi_sector}    & sector (A)  & - & \multicolumn{1}{c}{\begin{tabular}[c]{@{}c@{}}slab cache;\\driver-specific \CheckmarkBold\end{tabular}} \\


\rowcolor{mygray}
\code{ext4_extent->ee_block}  & sector   & - &   \XSolidBrush  \\

\code{extent_status->es_pblk} & sector & - & \code{ext4_es_cachep} \CheckmarkBold  \\

\rowcolor{mygray}
\code{extent_map->block_start}  & sector  &  - & \code{extent_map_cache}     \CheckmarkBold  \\

\code{extent_state->start}  & sector  & - & \XSolidBrush  \\




\rowcolor{mygray}
\code{buffer_head->b_blocknr}    & sector  & - & bh\_cachep   \CheckmarkBold  \\



\code{xfs_buf_map->bm_bn} & sector & - & \XSolidBrush  \\

\rowcolor{mygray}
\code{request->__sector} & sector  & - & kmalloc; driver-specific \CheckmarkBold  \\

\code{nvme_request->result} & sector & -  & kmalloc; driver-specific \CheckmarkBold  \\

\rowcolor{mygray}
\multicolumn{1}{c|}{\begin{tabular}[c]{@{}c@{}}\code{ext4_inode_info->}\\\code{vfs_inode->i_mapping}\end{tabular}} & \multicolumn{1}{c|}{\begin{tabular}[c]{@{}c@{}}page cache\\tree\end{tabular}} & -  & FS-specific  \CheckmarkBold  \\

\code{file->f_mapping} & \multicolumn{1}{c|}{\begin{tabular}[c]{@{}c@{}}page cache\\tree (A)\end{tabular}} &  -  & filp\_cachep  \CheckmarkBold\\


\rowcolor{mygray}
\code{pipe_buffer->page} & page cache   &  - & variable size 
\CheckmarkBold  \\

\multicolumn{1}{c|}{\begin{tabular}[c]{@{}c@{}}\code{address_space->i_pages}\\\code{(xarray)}\end{tabular}}  & page cache  & -  & radix\_tree\_node\_cachep \CheckmarkBold   \\

\rowcolor{mygray}
\code{bio_vec->bv_page*}    & page cache  &  - & \XSolidBrush    \\

\bottomrule[0.1pt]
\bottomrule[0.1pt]

\end{tabular}}
\vspace{-10pt}
\end{table}

\begin{table}[t]
\centering
\renewcommand{\arraystretch}{1.1}
\caption{Identified and verified critical FSKOs in the stack.}
\vspace{-5pt}
\label{tab:identified-stack}
\resizebox{1\linewidth}{!}{
\begin{tabular}{c|c|c|c}
\toprule[0.1pt]
\toprule[0.1pt]
 Data   &  Category     &  Target value       & Verified \\ \hline

\rowcolor{mygray}

\code{iov_iter->data_source}   & operation   & \code{WRITE}   &  \CheckmarkBold \\

\code{iomap_iter->flags} (not v5.14)   & operation   & \code{IOMAP_WRITE}   &  \CheckmarkBold \\

\rowcolor{mygray}
\multicolumn{1}{c|}{\begin{tabular}[c]{@{}c@{}}\code{blk_mq_alloc_data}\\\code{->cmd_flags}\end{tabular}} & operation  &  \code{REQ_OP_WRITE}   &  \CheckmarkBold   \\

\code{vm_fault->flags}  & COW  & \code{FAULT_FLAG_WRITE}    &   \CheckmarkBold  \\

\rowcolor{mygray}
\code{iomap->addr}        & sector   &  sector number  &      \CheckmarkBold  \\

\code{ext4_map_blocks->m_pblk}  & sector  &  sector number  &    \CheckmarkBold  \\

\rowcolor{mygray}
\code{xfs_bmbt_irec->br_startblock} & sector &  sector number  & \CheckmarkBold  \\

\code{erofs_map_blocks->m_pa} & sector &  sector number  & \CheckmarkBold  \\



\bottomrule[0.1pt]
\bottomrule[0.1pt]

\end{tabular}}
\vspace{-10pt}
\end{table}


\begin{table}[t]
\centering
\renewcommand{\arraystretch}{1.1}
\caption{The bridge objects facilitating page-UAF strategy. }
\vspace{-5pt}
\label{tab:identified-page}
\resizebox{1\linewidth}{!}{
\begin{tabular}{c|c|cccc}
\toprule[0.1pt]
\toprule[0.1pt]
 Data      & Memory Cache  & \multicolumn{2}{c}{Read/Write} \\ \hline

\multicolumn{1}{c|}{\begin{tabular}[c]{@{}c@{}}\code{address_space->i_pages}\\\code{(xarray)}\end{tabular}}    & radix\_tree\_node\_cachep   & $\divideontimes$  & $\bigstar$ \\

\rowcolor{mygray}
\code{pipe_buffer->page*}    & variable size   & $\divideontimes$   & $\bigstar$ \\

\code{bio_vec->bv_page*}       &  variable size   & $\divideontimes$   & $\bigstar$ \\

\rowcolor{mygray}
\code{mptcp_data_frag->page*} & page   &  & $\bigstar$  \\

\code{sock->sk_frag.page*} & \multicolumn{1}{c|}{\begin{tabular}[c]{@{}c@{}}slab cache/kmalloc;\\net-specific\end{tabular}} &  & $\bigstar$  \\


\rowcolor{mygray}
 \code{wait_page_queue->page*}(not v6.6)    & variable size   & $\divideontimes$   & $\bigstar$ \\

\code{xfrm_state->xfrag->page*}   & kmalloc-1k   & $\divideontimes$   & $\bigstar$ \\

\rowcolor{mygray}
\code{pipe_inode_info->tmp_page*}   & kmalloc-192  & $\divideontimes$   & $\bigstar$ \\
 \code{lbuf->l_page*}    & kmalloc-128  & $\divideontimes$   & $\bigstar$ \\

 \rowcolor{mygray}
 \code{skb_shared_info->frags->bv_page*}    & variable size   & $\divideontimes$   & $\bigstar$ \\


\code{cifs_writedata->pages**}(not v6.6)   & variable size      &  & $\bigstar$  \\

\rowcolor{mygray}
\code{cifs_readdata->pages**}(not v6.6)   & variable size      & $\divideontimes$ &   \\

\code{fuse_args_pages->pages**}  &variable size  &  & $\bigstar$ \\

\rowcolor{mygray}
\code{ceph_sync_write() **} &variable size  &  & $\bigstar$  \\

\code{ceph_sync_read() **} &variable size  &  $\divideontimes$ &   \\

\rowcolor{mygray}
\code{process_vm_rw_core() **} & variable size  & $\divideontimes$ &  $\bigstar$ \\


\multicolumn{1}{c|}{\begin{tabular}[c]{@{}c@{}}\code{orangefs_bufmap_desc}\\\code{->page_array**}\end{tabular}} & variable size    & $\divideontimes$ & $\bigstar$  \\

\rowcolor{mygray}
\code{vm_struct->pages**}   & variable size     & $\divideontimes$ &  \\

\code{agp_memory->pages**}  &variable size  &  & $\bigstar$ \\

\rowcolor{mygray}
\code{ttm_tt->pages**}  &variable size   &  & $\bigstar$ \\

 \code{Sg_scatter_hold->pages**}    & variable size    & $\divideontimes$   &  \\

\rowcolor{mygray}
 \code{st_buffer->reserved_pages**}    & variable size   & $\divideontimes$   & $\bigstar$ \\


\code{configfs_buffer->page}  &  kmalloc-128    & $\divideontimes$ & $\bigstar$  \\


\bottomrule[0.1pt]
\bottomrule[0.1pt]

\end{tabular}}

\begin{tablenotes} \scriptsize
    \item [3]{$\divideontimes$: the pointer can read information of the page content.}
    \item [3]{$\bigstar$: the pointer can write the page content through \code{page} pointer.}
\end{tablenotes}
\vspace{-15pt}
\end{table}

\subsection{Exploitable Objects}
\label{sec:exploitable-objects}

Overall, as shown in~\cref{tab:count}, our static analysis found 282 key fields within 149 structure objects for two versions. Then, 134 fields are verified dynamically with test cases, out of which 23 fields belonging to 17 FSKOs are confirmed to lead to privilege escalation. Note that all these 23 effective FSKO fields exist in both v5.14 and v6.6 versions, and we list all these 23 in \cref{tab:identified-flag} as well as a few other seemingly promising but did not get confirmed successfully. If a field in a table entry has no additional annotation in \cref{tab:identified-flag}, \cref{tab:identified-stack}, and \cref{tab:identified-page}, it means the field exists in both versions.
Note that we have deduplicated the objects by counting file-system-specific objects and fields only once, e.g., \code{ext2_inode} and \code{ext4_inode} are deduplicated. 
Even though not the focus of our research, we also find 8 extra objects in the stack.

%
The objects dynamically verified to be exploitable for privilege escalation are marked with \CheckmarkBold.
For the remaining ones, there are a few FSKO fields marked with \XSolidBrush that looked like promising fields (based on their names and usage patterns in the source code). However, when we verified them dynamically, overwriting their values did not lead to privilege escalation.
We listed a few cases where we do not have the right test cases to reach them dynamically (\textbf{---}). We also show unexploitable cases to demonstrate the necessity of dynamic verification because they have similar semantics to FSKOs.

Note that we exclude some redundant data from the results.
Specifically, the object \code{inode} is an embedded field and accessible through \code{ext4_inode_info->vfs_inode} of \code{inode} type. Therefore, the allocation of \code{inode} goes with the allocation of the parent object, i.e., \code{ext4_inode_info}. Each file system implementation has its own parent object, like \code{ext2_inode_info}. \name{} identifies 48 parent objects which are not listed in the paper.
%



Now we explain columns in \cref{tab:identified-flag}.
We show all FSKO fields (including the struct name) in the first ``Data'' column.
The ``Category'' column lists the semantics of the FSKO fields with (A) denoting anchor objects.
The ``Target value'' column represents the constants for escalate privilege by corrupting an FSKO field with the target value.
Regarding the memory cache in the last column,
a few objects are allocated in different caches depending on where they are allocated (drivers or network protocols) denoted with suffix ``-specific''.
%
%
In addition, some objects are allocated as arrays with variable sizes, and we denote them as ``variable size”. 
Some objects are directly allocated via the page allocator instead of the slab/slub allocator, and we use ``page'' in the table. 


\PP{Flag objects.} The beginning of \cref{tab:identified-flag} presents 13 flags that result in privilege escalation. Again, we deduplicated FSKOs that are file system specific and use objects under \code{ext4} that represent others.
%
The COW in~\cref{tab:count} and~\cref{tab:identified-flag} means a special type of flag that controls Copy-On-Write (COW).
These flags are in conditional branches controlling whether to obtain the existing page cache or allocate a new page.
It is worth mentioning that \name{} also identifies other COW objects for disk operations (instead of only the page cache in memory). However, we find that a normal user is not permitted to trigger a disk COW of a read-only file; thus, such objects are not included in our results.
%

\PP{Reference objects.}  We list 10 objects for sector numbers and 4 objects representing the page-cache tree in \cref{tab:identified-flag}. Substituting them with the reference for a read-only file enables unprivileged users to write it. We also use ext4 as an example, ignoring similar objects in other file systems.

\PP{Additional stack variables} We also verified 8 FSKO fields are in the stack shown in~\cref{tab:identified-stack}, which may raise security concerns for stack-based corruption~\cite{wu2019kepler, cho2020exploiting}. 
We only find the newly-introduced object \code{iomap_iter} in v6.6 and do not find it in v5.14. The other objects exist in both versions and remain stable. This demonstrates that our identified objects are related to the basic file system's functions (read/write), which are the core semantics and do not change frequently. 
Besides, DirtyCOW is a logic bug that manipulates the flag \code{vm_fault->flags} in the stack to achieve root. We believe these objects may also be valuable in future research on stack-based vulnerabilities and logic bugs.

\PP{Bridge Objects in page UAF Strategy}
\cref{tab:identified-page} show 23 exploitable objects facilitating the page UAF strategy. 
%
%
All these objects can be corrupted to point to other pages, thus creating page UAFs when they are freed. Then, we observe that 16 objects can be used to read ($\divideontimes$), and 20 write ($\bigstar$) the page content through their inside pointer. 12 can achieve both reading and writing. 
In the first column in \cref{tab:identified-page}, \code{*} signifies the field being of type \code{page *}. 
%
As for \code{**}, it means the field is of type \code{page **}, i.e., a heap-allocated array of \code{page *}. 
These arrays are often variable in size, and their elements can be corrupted to construct page UAF.
There is one entry (the first) marked with \code{xarray}, which is a specific data structure that is similar to \code{page **}.
Finally, some entries in \cref{tab:identified-page} are listed with function names ending with \code{()} and \code{**} (in the first column). 
They represent cases where an array of \code{page} pointers are allocated 
on the heap with \code{kmalloc_array()} without being assigned to a field of any heap object (instead it is assigned to a stack variable). 
Nevertheless, the \code{page} pointer array can still be corrupted.

\PP{Comparison with KENALI} We run the open-source code in GitHub, which uses generic heuristics to identify as many non-control data as possible across the Linux kernel, regardless of their semantics and exploitability. 
The idea is to look at various that can affect the control flow of the program, where one path leads to an error code in return and the other does not. After running it on the same kernel version v5.14, we found 189 struct types, and only 10 of them belong to the file subsystem. 2 of them structs were also found, namely \code{inode} and \code{address_space}. 
We investigated the remaining 8, and none of them can lead to privilege escalation using the three types of semantics we considered.

\begin{table*}[t!]
	\centering
	\renewcommand{\arraystretch}{1.2}
	\caption{Exploitability demonstrated on 18 real-world vulnerabilities using different exploit strategies. \emptycirc{} means the exploits cannot bypass CFI or data protection; \halfcirc{} means the exploits can bypass CFI but fail if data protection is enforced; \fullcirc{} means the exploits can bypass both CFI and data protection.}
	\label{tab:cve-table}
    \small
	\setlength\tabcolsep{10pt}
 \resizebox{.9\linewidth}{!}{
		\begin{tabular}{lccccccccc}
\toprule[0.1pt]
\toprule[0.1pt]
 \multirow{2}{*}{\textbf{CVE ID}} & \multirow{2}{*}{\textbf{CVE Type}} & \multirow{2}{*}{\textbf{Strategy}} & \multicolumn{2}{c}{\textbf{Existing Exploit}}  & \multicolumn{2}{c}{\textbf{Exploit FSKO}}\\
\cline{4-7}			
 &&& \multicolumn{1}{c}{\begin{tabular}[c]{@{}c@{}}\textbf{Bypass CFI \&}\\\textbf{Data Protection}\end{tabular}} &  \multicolumn{1}{c}{\begin{tabular}[c]{@{}c@{}}\textbf{No Need}\\\textbf{InfoLeak}\end{tabular}} & \multicolumn{1}{c}{\begin{tabular}[c]{@{}c@{}}\textbf{Bypass CFI \&}\\\textbf{Data Protection}\end{tabular}} &  \multicolumn{1}{c}{\begin{tabular}[c]{@{}c@{}}\textbf{No Need}\\\textbf{InfoLeak}\end{tabular}} \\
\midrule
\rowcolor{mygray}
 CVE-2023-5345 & Use-After-Free & Page UAF & \halfcirc & \XSolidBrush & \fullcirc & \CheckmarkBold \\

 {CVE-2022-0995} & Out-Of-Bounds & Corrupt Flag, Page UAF & \emptycirc & \XSolidBrush & \fullcirc & \CheckmarkBold \\

\rowcolor{mygray}
 {CVE-2022-0185} & Out-Of-Bounds & Page UAF & \fullcirc & \XSolidBrush & \fullcirc & \CheckmarkBold \\

 CVE-2021-22555 & Out-Of-Bounds & Page UAF  & \fullcirc & \XSolidBrush & \fullcirc & \CheckmarkBold  \\

\rowcolor{mygray}
 CVE-2021-22600 &  Double-Free & Corrupt Ref, Page UAF & \fullcirc & \XSolidBrush & \fullcirc & \CheckmarkBold \\


 CVE-2022-27666 & Out-Of-Bounds &  Page UAF & \halfcirc & \XSolidBrush & \fullcirc & \XSolidBrush \\

\rowcolor{mygray}
 CVE-2022-25636 & Out-Of-Bounds & Corrupt Flag,Page UAF & \fullcirc & \XSolidBrush & \fullcirc & \CheckmarkBold \\
 CVE-2022-2639 & Out-Of-Bounds & Page UAF & \fullcirc & \XSolidBrush & \fullcirc & \CheckmarkBold \\

\rowcolor{mygray}
 CVE-2022-2588 & Use-After-Free & Corrupt Flag, Page UAF  & \halfcirc & \CheckmarkBold & \fullcirc & \CheckmarkBold \\

 CVE-2021-3492 & Double Free &  Corrupt Flag, Page UAF & \emptycirc & \XSolidBrush & \fullcirc & \CheckmarkBold \\

\rowcolor{mygray}
 CVE-2021-43267 & Out-Of-Bounds & Page UAF & \emptycirc & \XSolidBrush & \fullcirc & \CheckmarkBold \\

 CVE-2021-41073 & Use-After-Free & Corrupt Flag, Page UAF & \emptycirc & \XSolidBrush & \fullcirc & \CheckmarkBold \\

\rowcolor{mygray}
 CVE-2021-4154 & Use-After-Free & Corrupt Flag, Page UAF& \halfcirc & \CheckmarkBold & \fullcirc & \CheckmarkBold \\

 CVE-2021-42008 & Out-Of-Bounds &  Page UAF & \fullcirc & \XSolidBrush & \fullcirc & \CheckmarkBold \\

\rowcolor{mygray}
 CVE-2021-27365 & Out-Of-Bounds & Page UAF & \emptycirc & \XSolidBrush & \fullcirc & \CheckmarkBold \\

 CVE-2021-26708 & Use-After-Free &  Page UAF & \emptycirc & \XSolidBrush & \fullcirc & \XSolidBrush \\

\rowcolor{mygray}
 CVE-2021-23134 & Use-After-Free &  Corrupt Flag, Page UAF & \halfcirc & \XSolidBrush & \fullcirc & \CheckmarkBold \\

 CVE-2020-14386 & Out-Of-Bounds &  Page UAF & \emptycirc & \XSolidBrush & \fullcirc & \CheckmarkBold \\

\bottomrule[0.1pt]
\bottomrule[0.1pt]
\end{tabular}}
\vspace{-10pt}
\end{table*}

\subsection{Exploitability Evaluation}

Out of the 26 CVEs mentioned earlier, we confirmed that 18 of them are exploitable, by attempting to pair FSKOs with the vulnerability in the two dimensions: (1) write capabilities, and (2) slab cache requirements. 
%
The 10 CVEs are not suitable for FSKOs because they occur in specific subsystems that are not general to heap objects, e.g., eBPF subsystem.


The results are shown in \cref{tab:cve-table}.
We list the exploit strategies suitable for each CVE.
To further demonstrate the exploitability, we develop 10 end-to-end exploits where the underlying Linux kernel enables all the exploit mitigation mechanisms: CVE-2021-22600 (Double Free), CVE-2021-22555 (OOB), CVE-2022-0995 (OOB), and CVE-2022-0185 (OOB), CVE-2023-5345 (UAF). 
These CVEs cover all three kinds of vulnerabilities: OOB, UAF, and double-free.
For the remaining CVEs, we have determined that they can also be exploited similarly based on their write capability and slab cache requirement. However, writing them is time-consuming because of tasks such as creating stable heap feng shui. 


\PP{Corrupting flag objects} 
Corrupting flag objects is straightforward as long as the write capability is suitable for the specific flag field. 
For this exercise, we write an exploit using CVE-2022-0995 (an OOB vulnerability) to corrupt the \code{pipe_buffer->flags}. This vulnerability happens to have the right capability to write this flag.
This CVE provides a capability that sets an arbitrary bit to 1 based on the vulnerable object of size 0x38. After placing the FSKO adjacent to the vulnerable object, we can directly corrupt the flag.

%
%

\PP{Corrupting reference objects}
We write an exploit using CVE-2021-22600 to corrupt the reference objects with the target value after infoleak. 
%
Specifically, this double-free vulnerability enables an attack to read the value of a page cache pointer to a read-only file (by spraying \code{pipe_buffer} objects). Subsequently, we trigger another double free to replace a read-write file's page cache pointer (again in \code{pipe_buffer}) to have it point to the read-only page cache. 
Note that this exploit does require infoleak. Fortunately, we can construct page UAF to avoid the need for infoleak for this CVE.

\PP{Corrupting FSKOs via page UAF}
We developed 8 exploits against CVE-2023-5345 (one exploit), CVE-2022-0995 (two exploits), CVE-2022-0185 (three exploits), CVE-2021-22555 (one exploit), and CVE-2021-22600 (one exploit), to corrupt FSKO fields, \code{pipe_buffer->flags}, \code{file->f_mode}, and \code{file->f_mapping}. The exploits spray bridge objects to construct page UAF -- we specifically target \code{pipe_buffer->page} and \code{configfs_buffer->page}.
These exploits are generally easier and more stable because there is no need for infoleak or page-level feng shui.

\textit{Complexity Reduction.}
Given some exploits do not require KASLR bypassing (i.e., the ones that utilize the page-UAF strategy or directly overwrite flag), we aim to measure the exploit complexity in terms of lines of code (LoC). 
For the exploits to corrupt flags, our exploits achieve 20\% and 70\% reduction in exploit LoC compared to the publicly available exploits. The case with 70\% is where the CVE's capability allows corrupt flags at the correct offset (without requiring the page-level UAF first). Even with page-level UAF, we still managed to see a 20\% reduction in LoC. 
For the other exploits, we observe approximately a 30\% reduction in LoC compared to the original exploits. 
Besides, some of the original exploits are based on return-oriented programming (ROP) and require an extra offline step of gadget extraction. This step is not necessary in our exploits. In summary, the complexity reduction is mainly due to the absence of infoleak requirements and ROP (collect gadgets and get their offsets).

\textit{Stability and Usability.} 
The page-level-UAF-based exploits have a high success rate, 90\% - 100\%. We ran our exploit of five exploits of CVE-2022-0995 and CVE-2022-0185. Each exploit is run 10 times, and 9 or 10 out of 10 are successful without any crashes. This is attributed to the fact that the exploits have a built-in feedback mechanism (i.e., read of the physical page) that helps make sure the final write is performed on the right target. Using the feedback mechanism, we can restart the page-level UAF if the expected target is not detected.

\textit{Future-proof.}
We compare our exploits with the publicly available exploits in~\cref{tab:cve-table}. Note that a CVE can have multiple public exploits, and we select the most powerful one that can bypass the most protections.
Some exploits are rendered ineffective under CFI deployment, indicated by \emptycirc{} for those that cannot bypass CFI. 
Besides CFI, it is a trend that some critical non-control data will be protected, e.g., \code{cred}, \code{modprobe_path}, and page table are already protected in Android kernels~\cite{samsung1, samsung2}. 
This will make a number of exploits obsolete in the near future.
%
We mark existing exploits with \halfcirc{} if they choose to corrupt the non-control data that are protected today in Android kernels (because they are likely going to be protected in Linux kernels in the future).
We use \fullcirc{} to represent the exploits that bypass CFI and data protection.
Notably, we have verified that our exploits are able to succeed, even when the kernel enables the protection against DiryCred, i.e., providing isolation of \code{file} objects by separating them into high-privileged and low-privileged ones. The reason this defense is ineffective against our exploit strategy is that we do not rely on co-locating high-privileged and low-privileged objects. Instead, once a page is freed, we can spray all high-privileged objects, e.g., spraying \path{/etc/passwd} files, and modify the permission \code{file->f_mode}. 

To the best of our knowledge, only five open-source existing exploits corrupt unprotected non-control data. But they all require info leak \XSolidBrush; our exploits do not \CheckmarkBold.
Four of them corrupt the flag \code{pipe_buffer->flags}, which is found in our paper. One (CVE-2021-22600) uses the User-Space-Mapping-Attack (USMA) technique that requires the target code address through infoleak. However, it typically leaks a specific address, where attackers calculate the target address by adding an offset based on the leaked address. The kernel code may differ version by version, resulting in the offset not being the same and the need to adapt for different kernels.

\section{Discussion}


\PP{Ethical consideration}
This work does not find any new vulnerabilities but exploits only existing ones with newly identified data. All exploited CVEs are patched already. The system is kept safe as long as it contains no exposed memory corruption vulnerabilities.

\PP{Defense}
The newly identified objects in this work can enhance existing protection mechanisms, fostering further research.
Many protection mechanisms safeguard critical non-control objects, e.g., data integrity using shadow memory~\cite{song2016enforcing} and supervisor-based protection~\cite{samsung1}. Researchers propose various techniques to prevent specific vulnerabilities from being exploited, e.g., eBPF-based monitor~\cite{wang2023PET}, slab allocator redesign~\cite{MineSweeper, HeapExpo, MarkUs, Brian2023Preventing, Randomiz32:online}. 
To mitigate the page UAF strategy we proposed, we first observe that it works because the kernel allows mixes of two types of pages:
(1) those freely accessible to user-space, e.g., \code{pipe_buffer} pages where arbitrary read and write can be performed from the user-space;
(2) those that are used as slab caches. 
Our proposal is simple: isolate these two types of page groups so that they never overlap in a page UAF situation. 
In fact, the Linux kernel already has mechanisms for page isolation~\cite{migratetype}, where between 4 and 6 page groups are enabled (depending on the kernel config).
We can address this concern by adding another page group.
\section{Related Work}

\PP{Kernel Vulnerability Exploitation.} 
Control flow hijacking is a powerful exploit~\cite{cficarlini2015control}. 
FUZE~\cite{wu2018fuze} and KOOBE~\cite{chen2020koobe} utilize various techniques 
to automatically generate exploits of control flow hijacking, targeting UAF and OOB vulnerabilities, respectively. 
KEPLER~\cite{wu2019kepler} leverages kernel-user interactions to convert control-flow hijacking into stack overflow and ROP attacks. 
GREBE~\cite{lin2022grebe} and SyzScope~\cite{zou2022syzscope} investigates fuzzer-exposed bugs and reveals the transformation of low-risk bugs into high-risk ones, such as control flow hijacking for kernel exploitation. 
The prevalence of control flow hijacking and code reuse attacks~\cite{zeng2022playing,chen2020systematic} in real-world incidents has spurred the development of defense mechanisms~\cite{ge2016fine,yang2019arm,denis2020camouflage,yoo2022kernel,linuxaddcfi:online}, making control-data related exploitation more challenging.

\PP{Non-Control Data Attacks.}
DOP \cite{hu2016data} shows code execution can be achieved by only chaining data gadgets. 
However, it relies on arbitrary write primitives, which are not always available in the real world. 
AlphaEXP~\cite{AlphaEXP} identifies different sensitive-level objects but does not focus on privilege escalation.
%
VIPER \cite{ye2023viper} identifies critical variables in user space programs that determine to invoke security-related system calls. 
However, existing works on kernel space data-only attacks merely concentrate on a limited set of known objects. 
Dirty page table \cite{DirtyPag12:online} corrupts the page table entry to launch attacks. DirtyCred \cite{lin2022dirtycred} offers a new exploit strategy to grant high-privilege credentials to normal users. 
DirtyCOW \cite{DirtyCOW66:online} exploits a race condition bug that exists in Copy-on-Write. 
DirtyPipe \cite{Arinerro42:online} leverages the use-before-initialization flag of a pipe to write to the page cache, bypassing permission checks. 
Previous work \cite{chen2005non} shows that user identity, configuration, input, and decision-making data can be exploited for attacks. The variables controlling the Linux auditing framework, AppArmor, and NULL pointer dereference mitigation can be bypassed through data attacks \cite{xiao2015kernel}. Our work aims to discover more critical but unknown non-control data in the Linux kernel.



\PP{Protection of Critical Data.} 
Control-flow integrity mechanisms in the mainline Linux kernel and default enforcement in Android~\cite{androidcfi} make hijacking control flow increasingly difficult. Additionally, commercial kernels employ advanced techniques to protect commonly exploited non-control data. These defenses can safeguard specific data from corruption even if an attack achieves arbitrary kernel-memory-write primitives. For example, Android uses the hypervisor to monitor \code{cred} and the page table via real-time protection~\cite{samsung1, samsung2}.
%
Meanwhile, the Page Protection Layer~\cite{appleblackhat}, an Apple-introduced feature, protects certain parts of the kernel from itself and is used to safeguard critical data such as \code{cred} and page tables~\cite{siguza-ios-mitigation,appleoss24:online}.
The global variable \code{modprobe_path}~\cite{li2023hybrid} can be marked as read-only with the config introduced in Linux kernel 4.11 (\code{CONFIG_STATIC_USERMODEHELPER}).
Research studies have proposed other ideas, e.g., software-based shadow memory~\cite{song2016enforcing} and hardware-based Extended Page Tables~\cite{proskurin2020xmp} that target a wide range of non-control objects.
Unfortunately, their effectiveness highly depends on their ability to identify meaningful non-control data. For instance, since KENALI~\cite{song2016enforcing} failed to identify the majority of FSKOs that we reported, our objects would not be put into the protection domain. 

\section{Conclusion}
This paper investigates the exploitability of non-control data in the Linux file system.
We systematically summarize three types of FSKOs for privilege escalation with the help of automated analyses. We analyze their exploitability by understanding their requirements of pairing with vulnerabilities of different capabilities. 
Along the way, we develop a novel strategy that can indirectly read and write FSKOs with high reliability.
Finally, using the discovered FSKOs, we develop end-to-end exploits using 18 recent real-world CVEs.

\bibliographystyle{IEEEtran}  
\bibliography{refs}


 





\end{document}